\documentclass[a4paper, 12pt]{article}
\usepackage[utf8]{inputenc}
\usepackage[T1]{fontenc}
\usepackage[top=2.5cm,bottom=2cm,left=2.5cm,right=2cm,asymmetric,twoside,headsep = 6mm]{geometry}
\usepackage{mathptmx} 
\usepackage{enumitem} 
\usepackage{graphicx}
\graphicspath{{../images/}} 
\usepackage{csquotes}  
\usepackage{amsmath}  
\usepackage{commath}  
\usepackage{amsthm} 
\usepackage{amssymb} 
\usepackage{subfigure} 
\usepackage[font=footnotesize]{caption} 
\usepackage{multirow} 
\usepackage{csquotes}  
\RequirePackage[authoryear]{natbib}

\usepackage{fancyhdr}
\def\R{\mathbb{R}}

\newtheorem{lemma}{Lemma}
\newtheorem{corollary}{Corollary}
\newtheorem{remark}{Remark}
\newtheorem{example}{Example}

\begin{document}
\thispagestyle{empty}
\vspace*{4em}
\noindent
\begin{minipage}{\textwidth}
\begin{flushleft}
\Huge
\textbf{Should we test the model assumptions before running a model-based test?}\\[3ex]
\Large 
\textbf{Iqbal Shamsudheen$^1$ and Christian Hennig$^2$}\\[2ex] 
\normalsize
\emph{$^{1}$ Department of Defence Science\\
  Faculty of Defence Science and Technology\\
  National Defence University of Malaysia, Kuala Lumpur, Malaysia\\
  E-mail: iqbal@upnm.edu.my\\
$^2$ Dipartimento di Scienze Statistiche, Universita di Bologna, Via delle belle Arti, 41, 40126 Bologna, Italy,\\
E-mail: christian.hennig@unibo.it}
\end{flushleft}
\end{minipage}
~\\[6ex]
\textbf{Abstract}\\
Statistical methods are based on model assumptions, and it is statistical folklore that a method's model assumptions should be checked before applying it. This can be formally done by running one or more misspecification tests of model assumptions before running a method that requires these assumptions; here we focus on model-based tests. A combined test procedure can be defined by specifying a protocol in which first model assumptions are tested and then, conditionally on the outcome, a test is run that requires or does not require the tested assumptions. Although such an approach is often taken in practice, much of the literature that investigated this is surprisingly critical of it.
Our aim is to explore conditions under which model checking is advisable or not advisable. For this, we review results regarding such ``combined procedures'' in the literature, we review and discuss controversial views on the role of model checking in statistics, and we present a general setup in which we can show that preliminary model checking is advantageous, which implies conditions for making model checking worthwhile.

\normalsize



\section{Introduction}\label{sintro}

Statistical methods are based on model assumptions, and it is statistical folklore that a method's model assumptions should be checked before applying it. Some authors believe that the invalidity of model assumptions and the failure to check them is at least partly to blame for what is currently discussed as ``replication crisis'' (\cite{Mayo18}), and indeed model checking is ignored in much applied work (\cite{KHLOCDKLPKL98,StZaMaPfUl07a,StZaMaPfUl07b,WJWGLMWHWWXH11,SriGow15,Nour-Eldein16}). Yet there is surprisingly little agreement in the literature about how to check the models. As will be seen later, several authors who investigated the statistical characteristics of running model checks before applying a model-based method come to negative conclusions. So is it sound advice to check model assumptions first? If so, are model misspecification tests a good tool for doing this? Here we present a comprehensive discussion of the issue, including a survey of the literature in which this is investigated. 

 

The amount of literature on certain specific problems that belong to this scope is quite large and we do not attempt to review it exhaustively. We restrict our focus to the problem of two-stage testing, i.e., hypothesis testing conditionally on the result of preliminary tests of model assumptions. More work exists on estimation after preliminary testing. For overviews see \cite{BanHan77,GilGil93,Chatfield95,Saleh06}. Almost all existing work focuses on analysing specific preliminary tests and specific conditional inference; here a more general view is provided.

\begin{figure}[t]
\caption{Diagram for a combined procedure involving misspecification (MS) testing for deciding which main test is used.}
\begin{center}
\includegraphics[width=0.7\textwidth]{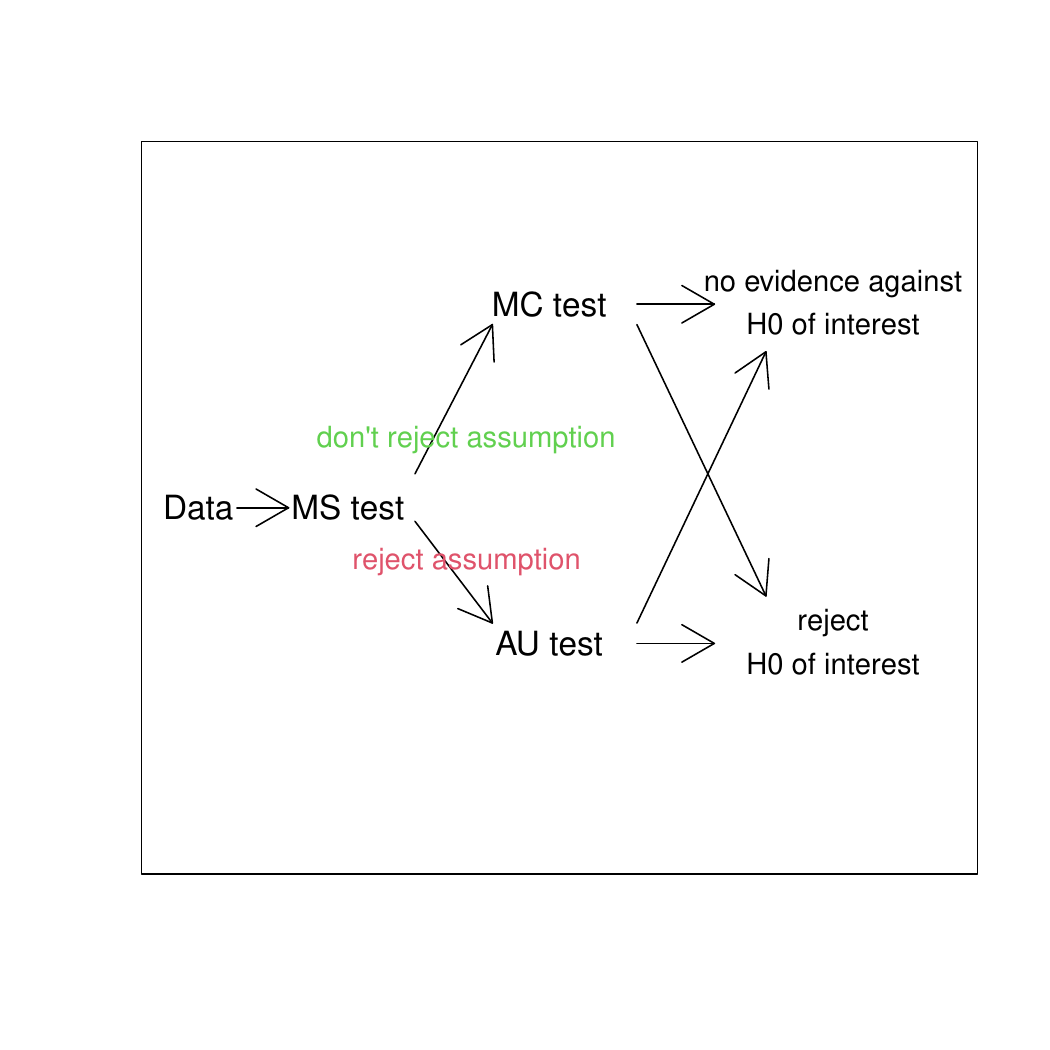}
\end{center}
\label{fig:scheme}
\end{figure}

To fix terminology, we assume a situation in which a researcher is interested in using a ``main test'' for testing a main hypothesis that is of substantial interest. There is a ``model-based constrained (MC) test'' involving certain model assumptions available for this. We will call ``misspecification (MS) test'' a test with the null hypothesis that a certain model assumption holds. We assume that this is not of primary interest, but rather only done in order to assess the validity of the model-based test, which is only carried out in case the MS test does not reject (or ``passes'') the model assumption. In case the MS test rejects the model assumption, there may or may not be an ``alternative unconstrained (AU) test'' that the researcher applies, which does not rely on the rejected model assumption, in order to test the main hypothesis. A ``combined procedure'' consists of the complete decision rule involving MS test, MC test, and AU test (if specified), see Figure \ref{fig:scheme}. More complex combined procedures can also be defined, in which more than one model assumption is tested, and more than one AU test could be considered, depending on which specific model assumption is rejected. 

As an example consider a situation in which a psychiatrist wants to find out whether a new therapy is better than a placebo based on continuous measurements of improvement on two groups of patients, one group treated with the new therapy, the other with the placebo. The researcher may want to apply a two-sample \textit{t}-test, which assumes normality (MC test). Normality can be tested, e.g., by a Shapiro-Wilk test (MS test) in both groups, and in case normality is rejected, the researcher may decide to apply a Wilcoxon-Mann-Whitney (WMW) rank test (AU test) that does not rely on normality. Such a procedure is for example applied in \cite{HolMye05,KoGiZeOrBa18}, and endorsed in some textbooks, see, e.g., Fig. 8.5 in \cite{DoWeCh04}. The following issues occur:
\begin{itemize}
\item The two-sample \textit{t}-test has further assumptions apart from normality, namely that the data within each group are independently identically generated (i.i.d.), the groups are independent, and the variances are homogeneous. There are also assumptions regarding external validity, such as the sample being representative for the population of interest, and the measurements being valid. Neither is the WMW test assumption free, even though it does not assume normality. Using only a single MS test, not all of these assumptions are checked, and all three tests, MS, MC, and AU may be invalidated, e.g., by problems with the i.i.d. assumption. Using more than one MS test for checking model assumptions (e.g., using the Shapiro-Wilk test as a first MS test of normality and the F-test as a second MS test of homogeneity of variances) before running the MC test may be recommended, and conditionally of the outcome of the MS tests, more than one AU test, see above. Investigating combined procedures involving a single MS and AU test, as is mostly done in the literature and in the present work as well, is a simplification, and can be seen as analysing the building blocks of a more complex strategy that may be applied in reality.
\item The two-sample \textit{t}-test tests the null hypothesis $H_0:\ \mu_1=\mu_2$ against $H_1:\ \mu_1\neq\mu_2$ (or larger, or smaller), where $\mu_1$ and $\mu_2$ are the means of the two normal distributions within the two groups. $H_0$ and $H_1$ are  defined within the normal model, and more generally, $H_0$ and $H_1$ of the MC test are defined within the assumed model. $H_0$ and $H_1$ tested by the AU test will not in general be equivalent, so there needs to be an explicit definition of the hypotheses tested by a procedure that depending on the result of the MS test will either run the MC or the AU test. In the example, in case that the variances are indeed homogeneous, the $H_0$ and $H_1$ tested by the \textit{t}-test are a special case of $H_0$ and $H_1$ tested by the WMW test, namely that the two within-groups distributions are equal ($H_0$) or that one is stochastically larger or smaller than the other ($H_1$). See \cite{FayPro10} for a discussion of different ``perspectives'' of what the WMW- and \textit{t}-test actually test. The combined procedure delivers a test of these more general $H_0$ and $H_1$, which sometimes may not be so easy to formalise. The key issue is how the scientific research question (whether the new therapy is equivalent to a placebo) translates into the specific model assumed by the MC test and the more general model assumed by the AU test.  
\end{itemize}    
The AU test may rely on fewer assumptions by being nonparametric as above, or by being based on a more general parametric model (such as involving an auto-regressive component in case of violation of independence). It does not necessarily have to be based on more general assumptions than the MC test, it could also for example apply the original model with a transformed variable. 

The central concept of the present work is that the question whether it is advisable to check the model assumptions first can be addressed by an analysis of the performance characteristics of combined procedures, i.e., type 1 and type 2 error probabilities (of rejecting a true null hypothesis, and of not rejecting the null in case the alternative is true; this is one minus the power) of the main test conditional on the result of the MS test, looking at both situations in which the model assumptions of the MC test are fulfilled, and where they are violated, potentially in various different ways, not necessarily even fulfilling the assumptions of the AU test. The issue of interest here is whether the performance characteristics of the combined procedure are good enough to recommend it, compared to running either the MC or the AU test unconditionally. We review existing results of this kind (all of which concern specific choices of MS, AU, and MC test), we review and discuss the concept in the context of general model assumption checking including some controversial views of it, we present a new theoretical result that has some implications for how to do MS testing, and we touch on some implications for practice. It is not the aim of the present work to make recommendations for specific problems; we rather outline a general attitude and guidelines for how to deal with model assumption testing. We will also comment on informal (visual) model checking. 

The binary nature of statistical tests has recently been controversially discussed (\cite{WasLaz16}). Our focus on type 1 and type 2 error probabilities is not meant to advocate simplistic ``accept/reject'' decisions in science, but rather to allow for a transparent analysis to improve the understanding of the procedures. Note though that the decision to proceed with a certain model-based method or not is essentially binary (potentially with more than two branches), so that binary decisions cannot be entirely avoided.      

It is well known, going back to \cite{Bancroft44}, that the performance characteristics of a combined procedure differ in general from the characteristics of the MC test run unconditionally, even if the model assumptions of the MC test are fulfilled. This can be seen as a special case of data-dependent analysis, called ``garden of forking paths'' by \cite{GelLok14}. They suggest that such analyses contribute to the fact that ``reported statistically significant claims in scientific publications are routinely mistaken'', particularly if the researchers choose freely, conditionally on test results, which of potentially many outcomes they report. The preliminary definition of combined procedures can serve as some kind of pre-registration to protect against leaving too many degrees of freedom to the researcher, and it allows to analyse the implications of what is done. 

We generally assume that the MS test is carried out on the same data as the main test. Some of the issues discussed here can be avoided by checking the model on independent data, however such data may not be available. See \cite{Chatfield95} for a discussion of obtaining the ``independent'' data by splitting the available data set, which will result in less data used for the main test, and therefore in loss of power. In any case it would leave open the question whether the data used for MS testing are really independent of the data used for the main test, and whether they do really follow the same model. If possible, this is however a valuable option.

The situation is confusing for the user in the sense that checking model assumptions is recommended in many places (e.g., \cite{Spanos99,Cox06,KaCaDaMeYuRe16}), but an exact formal specification of how to do this in any given situation is hardly ever given. Tests are routinely used in applied research to decide about model assumptions in all kinds of setups, often for deciding how to proceed further (e.g., \cite{GaBaVoBeSaAlHo02,MaFoGaRe09,HoKiJo12,Ravichandran12,Abdulhafedh17,WuHaTeGuAyLo19,HaSuScMi20}). Guidelines for statistical analysis plans in clinical trials advise to specify in advance how to test the model assumptions and what to do in case they are rejected (\cite{Gambleetal17}; \cite{VanGin16} make a similar case for social psychology), whereas \cite{WilAlb19} advise against specifying an algorithmic ``decision tree'' for model assumption checks when pre-registering studies. 

Regarding the setup above, different authors arrive at different conclusions. Whereas both \cite{RoGoKi12} and \cite{FayPro10} advised against normality testing, \cite{FayPro10} recommended the WMW test if in doubt, because the \textit{t}-test can strongly deteriorate in the presence of outliers and heavy tails. On the other hand, \cite{RoGoKi12} preferred the \textit{t}-test, based on simulations on various distributions with light tails. 

Users do not normally know, before seeing the data, which of these distributions is more relevant in their situation. Therefore surely there is a demand for a test or any formal rule to distinguish between situations in which the WMW test (or any other specific alternative to the \textit{t}-test) is better, and situations in which the \textit{t}-test is better, based on the observed data. But this problem is different from distinguishing normal from non-normal distributions, as which this is often framed, and which is what a normality test nominally addresses. In fact, the situation is ambiguous, and a combined procedure involving normality testing can well be competitive, see Example \ref{etwosam}.

\begin{example}\label{etwosam} For illustration consider the simulation of a two-sample situation with i.i.d. data within the two samples $i=1, 2$, and sample sizes $n_1=20,\ n_2=30$. We chose Welch's two-sample {\textit t}-test (\cite{Welch38}) as MC test, the WMW test as AU test, and the Shapiro-Wilk test (on aggregated residuals from the mean, at level 5\% as all involved tests) as MS test for defining a combined procedure. It might be argued that using nonparametric, resampling-based, or robust methods without MS testing could be preferable. Therefore
we also included a permutation test of the difference between group means (\cite{Pitman37}) run unconditionally.

Data have been generated from the normal distribution ${\cal N}(\mu_i,1)$, a
$t_3$-distribution shifted by $\mu_i$ (here the variance exists and the Central Limit Theorem (CLT) applies, so that the {\textit t}-test is asymptotically valid), an exponential distribution with mean $\mu_i$, and a skew normal distribution with skewness parameter $\alpha=3$ and the other parameters chosen so that the variance is 1 and the expected value is $\mu_i$. For the main null hypothesis we used $\mu_1=\mu_2=1$; for the main alternative $\mu_1=1,\ \mu_2=2$. We ran 100,000 replicates. The resulting type 1 and type 2 error probabilities are given in Table \ref{ttwosam1}. They show that what is preferable depends strongly on the underlying distribution. The WMW test is clearly superior to the {\textit t}-test for the $t_3$- and skew normal distribution, but inferior for the normal and exponential distribution. The Shapiro-Wilk test detects the $t_3$- and exponential distribution in most cases, so that the combined procedure's performance is between the t- and WMW test for these, but very close to the WMW test, which is good in the $t_3$-case, but not so good in the exponential case. The skew normal is detected in about 39\% of replications. The combined procedure is excellent (much better than WMW) regarding the type 2 error of the normal and the skew normal distribution, but comes with a slightly anti-conservative type 1 error.

The permutation test does not offer general protection either; it does not improve on the {\textit t}-test for $t_3$ and falls clearly behind it for the exponential case.  

Based on the 100,000 simulation runs, $p$-values for one-sided binomial tests of the true type 1 error probability being 0.05 are between 0.01 and 0.05 for the combined procedure in the normal and $t_3$-case, and $<0.001$ in the skew normal case; differences between powers interpreted above are all highly significant. Combined procedure and permutation test do not have a significantly better power than the {\textit t}-test for the normal distribution, and neither does the combined procedure compared to the WMW test for the exponential distribution.   

\begin{table}
\begin{tabular}{r|r|l|l} 
Procedure & Distribution & Type 1 error prob. & Type 2 error prob.\\
\hline
Welch's {\textit t} & normal & .0498 & .0785\\
WMW       & normal & .0475 & .0924\\
Combined  & normal & .0512 & .0782\\
Permutation & normal & .0487 & .0778\\
\hline
Welch's {\textit t} & $t_3$ & .0453 & .4127\\
WMW       & $t_3$ & .0482 & .2585\\
Combined  & $t_3$ & .0515 & .2700\\
Permutation & $t_3$ & .0450 & .4142\\
\hline
Welch's {\textit t} & exponential & .0474 & .3389\\
WMW       & exponential & .0471 & .4853\\
Combined  & exponential & .0476 & .4849\\
Permutation & exponential & .0455 & .4472\\
\hline
Welch's {\textit t} & skew normal & .0493 & .0868\\
WMW       & skew normal & .0481 & .0778\\
Combined  & skew normal & .0531 & .0735\\
Permutation & skew normal & .0478 & .0777
\end{tabular}
\caption{Simulated error probabilities of four two-sample test procedures under four different distributions. The type 1 error probabilities have been generated using expectation $\mu=1$ in both samples, the type 2 error probabilities have been generated using $\mu_1=1,\ \mu_2=2$ in the two samples.
\label{ttwosam1}
}
\end{table}

The example shows that 
\begin{itemize}
\item even if model assumptions are clearly violated, an MC test may be better than the AU test or the combined procedure (exponential distribution);
\item often (normal, $t_3$, skew normal) the combined procedure is beneficial in the sense that its type 2 error probability is close to the better one of the MC and AU test, but this does not always work (exponential);
\item the combined procedure may even reach a better power than both the MC and AU test (skew normal), but it may also happen that its type 1 error probability is higher than the nominal 0.05 (normal, $t_3$, skew normal) - this is different from the MC and AU test, which both respect the nominal level throughout this study;
\item the ``distribution free'' methods WMW and permutation test do not perform uniformly well over all distributions. 
\end{itemize}
Depending on what deviation from normality seems more relevant in a given situation (without knowing the true underlying distribution), and to what extent the weak anti-conservativity of the combined procedure is a concern, a case could be made for any of the methods.

Obviously results depend on the specific parameter values and sample sizes chosen, and a comprehensive simulation study would vary these and possibly also look at situations in which there are two different distributions in the two samples. One could also define a more complex combined procedure involving an independence and homoscedasticity test and a set of several AU tests accounting for different detected deviations from the model assumptions.
\end{example}

\begin{remark} Obviously, increasing the test size, it is possible to create tests with arbitrarily large power. Therefore achieving larger power may not be seen as worthwhile if it comes to the price of a larger type 1 error probability. Then again, this may not be seen as a problem as long as tests are still conservative, i.e., the type 1 error probability is still not larger than the nominal level of the test. Three empirical type 1 error probabilities for the combined procedure in Example \ref{etwosam} are larger than the nominal level 0.05, but in two instances the difference is so small that it is only just about significant at 5\% level with 100,000 replicates. In reality any model assumption, e.g., independence, may be violated, see Section \ref{spersp}, so that theoretical guarantees are never fully reliable. Given that this is so, a user may find very weak anticonservativity acceptable if this leads to substantially better power (\cite{AlBoKa00a} present theory that in the situation of Section \ref{svar} involves bounding the type 1 error probability at $\alpha(1+\epsilon)$ with specified small $\epsilon$, where $\alpha$ is the nominal level). In Example \ref{etwosam}, this cannot be said for the combined procedure compared with WMW and permutation in the skew normal case, but it may look more attractive compared with WMW in the normal case, and with permutation in the $t_3$ case. In general both type 1 and type 2 error probabilities should be considered; we will not make assumptions here regarding the amount of tolerable anticonservativity, but rather note that this issue often plays a role when comparing test performances.
\end{remark}

Given the difficulty to define a convincing formal approach, it is not surprising that informal approaches for model checking are often used. Many researchers do informal model checking (e.g., visual, such as looking at boxplots for diagnosing skewness and outliers, or using regression residual plots to diagnose heteroscedasticity or nonlinearity), and they may only decide how to proceed knowing the outcome of the model check (be it formal or informal), rather than using a combined procedure that was well defined in advance. In fact, searching the web for terms such as ``checking model assumptions'' finds far more recommendations to use graphical model assessment than formal MS tests. An obvious advantage of such an approach is that the researcher can see more specifically suspicious features of the data, often suggesting ways to deal with them such as transformations.
This may work well, however it depends on the researcher who may not necessarily be competent enough to do this better than a formal procedure, and it has the big disadvantage that it cannot be formally investigated and replicated by other researchers, which would certainly be desirable. See \cite{SchFin18} for a critical assessment of using ad hoc transformations in regression for improving normality.
Only if the way how the researcher makes a visual decision is formalised, this can be analysed as another combined procedure. 
 

Apart from the example, we will consider specific investigations only as far as already covered in the literature. 
In Section \ref{spersp} we present our general perspective of model assumption checking. 
Section \ref{scomb} formally introduces a combined procedure in which an MS test is used to decide between an MC and an AU main test. Section \ref{scontro} reviews the controversial discussion of the role of model checking and testing in statistics. Section \ref{sspectest} runs through the literature that investigated the impact of misspecification testing and the performance of combined procedures in various scenarios. In Section \ref{stheory} we present a new result that formalises a situation in which a combined procedure can be better than both the MC and the AU test. Section \ref{sconc} concludes the paper and gives some conditions for model assumption checking to work well, which may stimulate further research and development of model checking procedures that are better than existing ones at finding those issues with the model assumptions that matter.
 
\section{A general perspective on model assumption checking} \label{spersp}
Our view of model assumptions and assumption checking is based on the idea that models are thought constructs that necessarily deviate from reality but can be helpful devices to understand it (\cite{Hennig10}, with elaboration for frequentist and Bayesian probability models in Section 5 of \cite{GelHen17}). Models can be used to show that certain procedures are good or even optimal in a certain sense, such as the Neyman-Pearson Lemma for tests; the WMW-test is never (to our knowledge) justified by an optimality result but rather by results warranting the validity of the level and the unbiasedness against certain alternatives, see \cite{FayPro10}; unbiasedness means that the rejection probability is always $\le\alpha$ (nominal level) throughout the $H_0$, and $\ge\alpha$ throughout the $H_1$. The term ``model assumption'' generally refers to the existence of such results, meaning that a method has a certain guaranteed quality if the model assumptions hold. 

We do not think that it is ever appropriate to state that any model is ``really true'' or any model assumption ``really holds''. The best that can be said is that it may be appropriate and useful to treat reality as if a certain model were true, acknowledging that this is always an idealisation. A test generally checks whether observed data are compatible with a model in a certain respect, which is defined by the test statistic. All data are compatible with many models; there are always alternatives to any assumed model that cannot be ruled out by the data, such as non-identical distributions that allow a different parameter choice for each observation or certain dependence structures (\cite{Hennig23}). Starting from the classical work of \cite{BahSav56}, there are results on the impossibility to identify from arbitrarily large data certain features of general families of distributions such as their means, or bounds on the density (\cite{Donoho88}). This means that it is ultimately impossible to make sure that model assumptions precisely or even only approximately hold.  

In order to increase our understanding of the performance of a statistical procedure, it is instructive to not only look at its results in situations in which the model assumptions are fulfilled, but also to explore it on models for which they are violated, but chosen so that if they were true, applying the procedure of interest still seems realistic.   
Such an approach is taken in Example \ref{etwosam}, the literature discussed in Section \ref{sspectest}, as well as in much literature on robust statistics, the latter mostly interested in worst case considerations (e.g., \cite{HaRoRoSt86}). According to \cite{Tukey97}, confronting a number of procedures with a ``bouquet of challenges'', i.e., data generated in various ways as in Example \ref{etwosam}, is a key guideline for data analysis in the absence of knowledge of a ``true model''.  

The problem that a procedure is meant to solve is often defined in terms of the assumed model. If other models are considered for data generation, an analogous problem has to be defined for those other models, which may not always be unique. A suitable way to think about this is that there is an informal scientific hypothesis and alternative of interest (such as ``no difference between treatments'' vs. ``treatment A is better'') that can be translated into various probability models, potentially in more than one way (e.g., ``treatment A is better'' may in a nonparametric setting translate into ``treatment A's distribution is stochastically larger'', or ``treatment A's distribution is a positive shift of treatment B's distribution'', or ``the expected outcome value of treatment A is larger''). Such a ``translation'' is required in order to investigate error probabilities on other than the originally assumed models, and to assess whether these are still satisfactory.

The problem of checking model assumptions is often wrongly framed as ``checking whether the model assumptions hold''. In reality they will not hold precisely anyway, but a method may still perform well in that case, and the model assumption may not even be required to hold ``approximately'' (see the exponential case in Example \ref{etwosam}). But there are certain violations of the model assumptions that have the potential to mislead the results in the sense of giving a wrong assessment of the underlying scientific hypothesis with high probability (e.g., for $t_3$ in Example \ref{etwosam} the probability is rather large that the t-test will find data from the $H_1$ compatible with the $H_0$ where other procedures can often show that they are not). ``Checking the model assumptions'' should rule such situations out as far as possible. This implies that model assumption checking needs to distinguish problematic violations from unproblematic ones, rather than distinguishing a true model from any wrong one. 
We think that some assumption checking does not work very well (see Section \ref{sspectest}) because it tries to solve the latter problem instead of the former.


Model assumption checking and choosing a subsequent method of inference conditionally on it, i.e., combined procedures, may help if done right, but may not help or even hurt if done wrong. Investigation of how well these procedures work in all kinds of relevant situations is therefore of interest. This is however hard, because the performance depends on all kinds of details, including the choice of MS, MC, and AU test, and particularly the models on which the procedures are assessed. Unfortunately, assuming that data dependent decisions are not made before the combined procedure is applied, the user may have little information about what distribution to expect, so that a wide range of possibilities is conceivable, and different authors may well come to different conclusions regarding the same problem (see the Introduction) based on different considered alternatives to the assumed model. This makes worst case considerations as in robust statistics attractive, but looking at a range of specific choices will give a more comprehensive picture. Given that certain violations of assumptions cannot be detected from the data alone, knowledge of the context (such as sampling schemes, measurement procedures, previous studies) should always be used to highlight potential issues on top of what can be diagnosed from the data.

\section{Combined procedures} \label{scomb}

The general setup is as follows. Given is a statistical model 
\begin{equation*}
	M_\Theta=\{P_\theta, \theta\in\Theta\} \subset M,
\end{equation*}
where $P_\theta, \theta\in\Theta$ are distributions over a space of interest, indexed by a parameter $\theta$. 
$M_\Theta$ is written here as a parametric model, but we are not restrictive about the nature of $\Theta$. $M_\Theta$ may even be the set of all i.i.d. models for $n$ observations, in which case $\Theta$ would be very large. However, in the literature,  $M_\Theta$ is usually a standard parametric model with $\Theta\subseteq\R^m$ for some $m$. There is a model $M$ containing distributions that do not require one or more assumptions implied by the definition of $M_\Theta$, but for data from the same space.  

Given some data $z$, we want to test a parametric null hypothesis 
$\theta\in\Theta_0$, which has some suitably chosen ``extension'' 
$M^*\subset M$ so that $M^*\cap M_\Theta=M_{\Theta_0}$, against the alternative
$\theta\not\in\Theta_0$ corresponding to $M\setminus M^*$ in the bigger model. 
In some cases (for example when applying the original model to transformed 
variables) $M$ may not contain $M_\Theta$, and $M^*\subset M$ then needs to be 
some kind of ``translation'' of the research hypothesis $M_{\Theta_0}$ into $M$,
the choice of which should be context guided and may or may not be 
trivial (e.g., equal group means for normals may be chosen to 
translate into equal medians rather than means when looking at distributions 
that can produce gross outliers with a certain non-vanishing probability).

In the simplest case, there are three tests involved, namely the MS test $\Phi_{MS}$, the MC test $\Phi_{MC}$ and the AU test $\Phi_{AU}$.  Tests can take values 0 or 1, where 1 means rejection of the null hypothesis.
Let $\alpha_{MS}$ be the level of $\Phi_{MS}$, i.e., $Q(\Phi_{MS}(z)=1)\le  \alpha_{MS}$ for all 
$Q\in M_\Theta$. Let $\alpha$ be the level of the two main tests, i.e.,
$P_\theta(\Phi_{MC}(z)=1)\le  \alpha$ for all $P_\theta, \theta\in\Theta_0$ and
$Q(\Phi_{AU}(z)=1)\le  \alpha$ for all $Q\in M^*$. To keep things general, for 
now we do not assume that type 1 error probabilities are uniformly equal to 
$\alpha_{MS}$, $\alpha$, respectively, and neither do we assume tests to be 
unbiased (which may not be realistic considering a big nonparametric $M$).

The combined test is defined as
\begin{equation*}
\Phi_C(z)=\left\{
\begin{array}{ccc}
\Phi_{MC}(z) & \mbox{ if } & \Phi_{MS}(z)=0,\\
\Phi_{AU}(z) & \mbox{ if } & \Phi_{MS}(z)=1.
\end{array}
\right.
\end{equation*}
This allows to analyse the characteristics of $\Phi_C$, particularly its 
effective level (which is not guaranteed to be $\le\alpha$) and power under 
$P_\theta$, or under distributions from $M^*$ 
or $M\setminus M^*$. Theoretical results are often hard to obtain without making
restrictive assumptions, although some exist, see Sections \ref{svar} and 
\ref{sreg}. At the very least, simulations are possible 
picking specific $P_\theta$ or $Q\in M$, and in many cases results may 
generalise 
to some extent because of invariance properties of model and test. 

Also of potential interest are 
$P_\theta\left(\Phi_C(z)=1|\Phi_{MS}(z)=0\right)$, i.e.,
the type 1 error probability under $M_{\Theta_0}$ or the power under 
$M_{\Theta}$
in case the model was in fact passed by the MS test, 
$Q\left(\Phi_C(z)=1|\Phi_{MS}(z)=0\right)$ for $Q\in M\setminus M_{\Theta}$, 
i.e., the situation that the model $M_{\Theta}$ is in fact violated but was 
passed by the MS test, and whether $\Phi_C$ can compete with $\Phi_{AU}$ in case
that $\Phi_{MS}(z)=1$ ($M_{\Theta}$ rejected). These are investigated in some of
the literature, see below. 



\section{Controversial views of model checking} \label{scontro}
The necessity of model checking has been stressed by many statisticians for a long time, and this is what students of statistics are often taught. \cite{Fisher22} stated: 
\begin{quote}
For empirical as the specification of the hypothetical population may be, this empiricism is cleared of its dangers if we can apply a rigorous and objective test of the adequacy with which the proposed population represents the whole of the available facts. 
\end{quote}
\cite{Neyman52} outlined the construction of a mathematical model, and emphasised checking the assumptions of the model by observation. \cite{Pearson00} introduced the goodness of fit chi-square test, which was used by Fisher to test model assumptions. The term ``misspecification test'' was coined as late as \cite{Fisher61} for the selection of exogenous variables in economic models. \cite{Spanos99} used the term extensively. \cite{Box80} emphasised the use of both MS tests and informal model diagnostics for Bayesian modelling.
See \cite{Spanos18} for the history and an exhaustive discussion of the use of MS tests. 

At first sight, model checking seems essential for two reasons. Firstly, statistical methods that a practitioner may want to use are often justified by theoretical results that require model assumptions. Secondly, it is easy to construct examples for the breakdown of methods in case model assumptions are violated in critical ways (e.g., inference based on the arithmetic mean, optimal under the assumption of normality, will break down for data generated from a Cauchy distribution).

Regarding the foundations of statistics,
checking of the model assumptions plays a crucial role in \cite{Mayo18}'s philosophy of ``severe testing'', in which frequentist significance tests are portrayed as major tools for subjecting scientific hypotheses to tests that they could be expected to fail in case they were wrong; and evidence in favour of such hypotheses can only be claimed in case they survive such severe probing. Mayo acknowledged that significance tests can be misleading in case the model assumptions are violated, but model assumptions themselves can be tested, and this offers some protection. A problem with this is that to our knowledge there are no results regarding the severity of MS tests, meaning that it is unclear to what extent a non-rejection of model assumptions implies that they are indeed not violated in ways that endanger the validity of the main test.    

A problem with preliminary model checking is that the theory of the model-based methods usually relies on the implicit assumption that there is no data-dependent pre-selection or pre-processing. A check of the model assumptions is a form of pre-selection. This is largely ignored but occasionally mentioned in the literature. \cite{Bancroft44} was probably the first to show how this can bias a model-based method after model checking. \cite{Chatfield95} gives a more comprehensive discussion of the issue. \cite{Hennig07} coined the term ``goodness-of-fit paradox'' (from now on called ``misspecification paradox'' here) to emphasise that in case model assumptions hold, checking them in fact actively invalidates them. Assume that the original distribution of the data fulfills a certain model assumption. Given a probability $\alpha>0$ that the MS test rejects the model assumption if it holds, the conditional probability for rejection under passing the MS test is obviously $0<\alpha$, and therefore the conditional distribution must be different from the one originally assumed. It is this conditional distribution that eventually feeds the model-based method that a user wants to apply. 

How big a problem is the misspecification paradox, and more generally the fact that MS tests cannot technically ensure the validity of the model assumptions? Some authors (e.g. \cite{Shuster05} and \cite{WelHin07}, who recommend considering model assumptions pre-data, if possible using earlier data and pilot studies) are generally dismissive of preliminary MS testing. On the other hand, \cite{Spanos10} argues that it is not a problem at all, because the MS test and the main test ``\textit{pose very different questions to data}''. The MS test tests whether the data ``\textit{constitute a truly typical realisation of the stochastic mechanism described by the model}''. He argues that therefore model checking and model-based testing can be considered separately; model checking is about making sure that the model is ``\textit{valid for the data}'' (\cite{Spanos18}), and if it is, it is appropriate to go on with the model-based analysis.

The point of view taken here, as in \cite{Chatfield95,Hennig07}, and elsewhere in the literature reviewed below, is different: We should analyse the characteristics of what is actually done. In case the model-based (MC) test is only applied  if the model is not rejected, the behaviour of the MC test should be analysed conditionally on data not being rejected by the MS test. We do not think that the misspecification paradox automatically implies that combined procedures are invalid; as argued in Section \ref{spersp} we do not believe that the model assumptions are true in reality anyway, and a combined procedure is worthwhile if it has good performance characteristics regarding the underlying scientific hypothesis, which may have formalisations regarding both the assumed model and the usually more general model employed by the AU test.  

If the distribution of the test statistic is independent of the outcome of the MS test, formally the misspecification paradox still holds, but it is statistically irrelevant. Conditioning on the result of the MS test will not affect the statistical characteristics of the MC test. An example for this is a MS test based on studentised residuals and a main test based on the minimal sufficient statistic of a normal distribution (\cite{Spanos10}). 
More generally it can be expected that if what the MS test does is at most very weakly stochastically connected to the main test (i.e., if in Spanos's terms they indeed ``pose very different questions to the data''), differences between the conditional and the unconditional behaviour of the MC test should be small. This can be investigated individually for every combination of MS test and main test, and there is no guarantee that the result will always be that the difference is negligible, but in many cases this will be the case.   

This alone does not imply that MS testing is beneficial, though. Also the case of violated model assumptions needs to be studied. In this case the conditional distribution of the AU test given that the MS test rejects the model assumption, and of the MC test given that the MS test does not reject, do matter.


Some kinds of visual informal model checking can be thought of as useful in a relatively safe manner if they lead to model rejections only in case of strikingly obvious assumption violations that are known to have an impact (which can be more precisely assessed looking at the data in a more holistic manner than a formal test can). In this case the probability to reject a true model can be suspected to be very close to zero, in turn not incurring much ``pretest bias''. But this relies on the individual researcher and their competence to recognise a violation of the model assumptions that matters. On the other hand, starting from around 1980 there were attempts to automatise data analysis, including deciding about model assumptions, in order to allow for the absence of expert statisticians, but also for improving reproducibility (\cite{Hand84,ShySpi14}).

A view opposite to Spanos's one, namely that model checking and inference given a parametric model should not be separated, but rather that the problems of finding an appropriate  distributional ``shape'' and parameter values compatible with the data should be treated in a fully integrated fashion, can also be found in the literature (\cite{Easterling76,Draper95,Davies14}). \cite{Davies14} argues that there is no essential difference between fitting a distributional shape, an (in)dependence structure, and estimating a location (which is usually formalised as parameter of a parametric model, but could as well be defined as a nonparametric functional).  

Bayesian statistics allows for an integrated treatment by putting prior probabilities on different candidate models, and averaging their contributions. Robust and nonparametric procedures may be seen as alternatives not only in case model assumptions of model-based procedures are violated; they have also been recommended for unconditional use (\cite{HaRoRoSt86,HolSet01}), making prior model checking supposedly superfluous. Example \ref{etwosam} however shows that ``model-free'' procedures without MS testing will not always perform very well. 
All these approaches still make assumptions; the Bayesian approach assumes that prior distribution and likelihood are correctly specified, robust and nonparametric methods still assume data to be i.i.d., or make other structural assumptions violation of which may mislead the inference (see \cite{WatHol16} for a Bayesian view of model misspecification and robustness). So the checking of assumptions issue does not easily go away, unless it is claimed (as some subjectivist Bayesians do) that such assumptions are subjective assessments and cannot be checked against data; for a contrary point of view see \cite{GelSha13}. To our knowledge, however, there is hardly any literature assessing the performance of model checking combined in which the ``MC role'' is taken by  robust, nonparametric or Bayesian inference, but see \cite{Bickel15} for a combined procedure that involves model checking and robust Bayesian inference. 

Some authors in the econometric literature (\cite{HenDoo14,Spanos18}) prefer ``respecification'' of parametric models, i.e., setting up a new parametric model accounting for the detected violation of the model assumptions, to robust or nonparametric approaches in the case that model assumptions are rejected. In some situations the advantage of respecification is obvious, particularly where a specific parametric form of a model is required, for example for prediction and simulation. More generally, \cite{Spanos18} argues that the less restrictive assumptions of nonparametric or robust approaches such as moment conditions or smooth densities are often untestable, as opposed to the more specific assumptions of parametric models. But this seems unfair, because to the extent that violations from such assumptions cannot be detected for more general models, it cannot be detected that any parametric model holds either. Impossibility results such as in \cite{BahSav56} or \cite{Donoho88} imply that distributions violating conditions such as bounded means, higher order moments, or existing densities are indistinguishably close to any parametric distribution. 
Ultimately Spanos is right that nonparametric and robust methods are not 100\% safe either, but they will often work under a wider range of distributions than a parametric model; e.g., classical robust estimation does safeguard against mixture distributions of the type $(1-\epsilon){\cal N}+\epsilon Q$, where ${\cal N}$ refers to a normal distribution, $Q$ to any distribution, $0<\epsilon$ small enough, which can have arbitrary or non-existing means and cannot be distinguished from a normal distribution with large probability for a given fixed sample size and $\epsilon$ small enough. Ultimately parametric respecification can be useful and can be successful in some cases such as sufficiently regular violations of independence where robust and nonparametric tools are lacking. Regarding the setup of interest here, the AU test can legitimately be derived from a parametric respecification of the model. When it comes to general applicability, in our view the cited authors seem too optimistic regarding whether a respecified model that can be confirmed by MS testing of all assumptions (as required by Spanos) to be reasonably valid can always or often be found. Cited results in Section \ref{sspectest} suggest in particular that situations in which a violated model assumption is not detected by the MS test for testing that very assumption can harm the performance of the MC test in a combined procedure. Furthermore, a respecification procedure as implied by Spanos including testing all relevant assumptions is to our knowledge not yet fully formalised and will be hard to formalise given the complexity of the problem, so that currently its performance characteristics in various possible situations cannot be investigated systematically. However, Spanos makes a valid point regarding the dangers of testing certain model assumptions in isolation, as MS as well as AU tests come with their own assumptions that may themselves be violated.  

Another potential objection to model assumption checking is that, in the famous words of George Box, ``all models are wrong but some are useful''. It may be argued that model assumption checking is pointless, because we know anyway that model assumptions will be violated in reality in one way or another (e.g., it makes some sense to hold that in the real world no two events can ever be truly independent, and continuous distributions are obviously not ``true'' as models for data that are discrete because of the limited precision of all human measurement). This has been used as argument against any form of model-based frequentist inference, particularly by subjectivist Bayesians (e.g., \cite{deFinetti74}'s famous ``probability does not exist''). \cite{Mayo18} however argues that ``all models are wrong'' on its own is a triviality that does not preclude a successful use of models, and that it is still important and meaningful to test whether models are adequately capturing the aspect of reality of interest in the inquiry. According to Section \ref{spersp}, it is at least worthwhile to check whether the data are incompatible with the model in ways that will mislead the desired model-based inference, which can happen in a Bayesian setting just as well. This does not require models to be ``true''. 


\section{Results for some specific test problems}\label{sspectest}
In this Section we will review and bring together results from the literature investigating the performance characteristics of combined procedures. Our focus is not on the detailed recommendations, but on general conditions under which combined procedures have been compared to unconditional use of MC or AU test, and have been found superior or inferior. Published results almost exclusively concern simple standard problems.
\subsection{The problem of whether to pool variances} \label{svar}
Historically the first problem for which preliminary MS testing and combined procedures were investigated was whether to test the equal variances assumption for comparing the means of two samples. Until now this is the problem for which most work investigating combined procedures exists. Let $X_1, X_2, ..., X_n$ be distributed i.i.d. according to $P_{\mu_1,\sigma_1^2}$ and $Y_1, Y_2, ..., Y_n$ be distributed i.i.d. according to $P_{\mu_2,\sigma_2^2}$, where $P_{\mu,\sigma^2}$ denotes a distribution with mean $\mu$ and variance $\sigma^2$. Most but not all literature consider $P_{\mu,\sigma^2}={\cal N}(\mu,\sigma^2)$, the normal distribution. If $\sigma_1^2=\sigma_2^2$, the standard two-sample 
\textit{t}-test using a pooled variance estimator from both samples (MC test) is optimal. 
For $\sigma_1^2\neq\sigma_2^2$ Welch's approximate \textit{t}-test with adjusted degrees of freedom depending on the two individual variances (AU test) is often recommended, see \cite{Welch38,Satterthwaite46,Welch47}. 

The equal variances assumption is
the historical starting point for investigations of the impact of MS testing. Early authors beginning from \cite{Bancroft44} did not frame the problem in terms of ``making sure that model assumptions are fulfilled'', but rather asked, in a pragmatic manner, under what circumstances pooling variances is advantageous. If the two variances are in fact equal or very similar, it is better to use all observations for estimating a single variance hopefully precisely, whereas if the two variances are very different, the use of a pooled variance will give a biased assessment of the variation of the means and their difference.   

It has been demonstrated that the two sample \textit{t}-test is very robust against violations of equality of variances when sample sizes are equal as shown by \cite{Hsu38,Scheffe70,PoYeOw82,Zimmerman06}. When both variances and sample sizes are unequal, the probability of the Type-I error exceeds the nominal significance level if the larger variance is associated with the smaller sample size and vice versa (\cite{Zimmerman06,WieAle07,Moder10}), which is amended by Welch's \textit{t}-test. \cite{BanHan77} published a bibliography of the considerable amount of literature on that problem available already at that time. One reason for the popularity of the variance pooling problem in early work is that, as long as normality is assumed, only the ratio of the variances needs to be varied to cover the case of violated model assumptions, which makes it easier to achieve theoretical results without computer-intensive simulations. 

Work that investigated sizes and/or power of combined procedures involving an MS test for variance equality for a main test of the equality of means, theoretically or by simulation,  comprises \cite{GurMcC62,Bancroft64,Gans81,MoStWa89,GupSri93,MosSte92,AlBoKa00a,Zimmerman14}. General findings are that the combined procedure can achieve a competitive performance regarding power and size beating Welch's \textit{t}-test
only in small subspaces of the parameter space with specific sample sizes. None of these authors recommends it for default use; \cite{MosSte92} recommended to never test the equal variances assumption. Often the unconditional Welch's \textit{t}-test is recommended, which is only ever beaten by a very small margin by the MC test or the combined procedure in specific situations. Occasionally the MC test is recommended conditionally on sample sizes being very similar.

\cite{MarMar90} hinted at what the problem with the combined procedure is. They evaluated the \textit{F}-test as MS test of homogeneity of variances for detecting deviations from variance equality that are known to matter for the standard \textit{t}-test by simulations, and showed that the \textit{F}-test is ineffective at finding these. Like \cite{Gans81}, they also involved non-normal distributions in their comparisons. This did not lead to substantially different recommendations.

\cite{AlBoKa00a} presented a second order asymptotic analysis of the combined procedure for pooling variances with the \textit{F}-test as MS test. They argue that this procedure can only achieve a better power than the MC test if the combined procedure also has a larger type 1 error probability. This means that there are only two possibilities for the combined procedure to improve upon the MC test. Either the combined procedure is anti-conservative, i.e., violates the desired test level, which would be deemed unacceptable in many applications, or the size of the MC test is smaller than the nominal level, which if its assumptions are not fulfilled is sometimes the case. \cite{AlBoKa00b} extend these results to the analysis of a more general problem for distributions $P_{\theta,\tau}$ from a parametric family with two parameters $\theta$ and $\tau$, where $\theta=0$ is the main null hypothesis of interest and the decision between an MC test assuming $\tau=0$ and an AU test without that assumption is made based on an MS test testing $\tau=0$. In the two-sample variance pooling problem, $\tau$ could be the logarithm of the ratio between the variances; a simpler example would be the choice between Gauss- and \textit{t}-test in the one-sample problem, where the MS test tests whether the variance is equal to a given fixed value. 
Once more, the combined procedure can only achieve better power at the price of a larger size, potentially being anti-conservative. Another key aspect is that the authors introduced a correlation parameter $\rho$ formalising the dependence between the MS-test and the main tests. In line with the discussion in Section \ref{scontro}, they state that for strong dependence preliminary testing is not sensible. 


\subsection{Tests of normality in the one-sample problem}\label{snorm1}

The simplest problem in which preliminary misspecification testing has been investigated is the problem of testing a hypothesis about the location of a sample. The standard model-based procedure for this is the one-sample Student's \textit{t}-test. It assumes the observations $X_1, X_2, ..., X_n$ to be i.i.d. normal. 
For non-normal distributions with existing variance the \textit{t}-test is asymptotically equivalent to the Gauss-test, which is asymptotically correct due to the CLT. The \textit{t}-test is therefore often branded robust against non-normality if the sample is not too small, see, e.g., \cite{Bartlett35,LehRom05}. An issue is that the quality of the asymptotic approximation does not only depend on $n$, but also on the underlying distributional shape, as the speed of improvement of the normal approximation is not uniform. Very skew distributions or extreme outliers can affect the power of the \textit{t}-test for large $n$, see \cite{Cressie80}. Another problem with the application of the CLT is the non-robustness of the mean as functional in the space of distributions (\cite{HaRoRoSt86}). Consider a situation in which the interest is in inference about $\mu$ assuming the underlying distribution as ${\cal N}(\mu,\sigma^2)$. If in fact the underlying distribution is $Q_{\mu.\sigma^2,\epsilon,x}=(1-\epsilon){\cal N}(\mu,\sigma^2)+\epsilon\delta_x$ with small $\epsilon$ and $x$ for away from $\mu$, $\delta_x$ being the Dirac measure in $x$, $EQ_{\mu.\sigma^2,\epsilon,x}$ may be arbitrarily far away from $\mu$ despite $d({\cal N}(\mu,\sigma^2), Q_{\mu.\sigma^2,\epsilon,x})$ small for most sensible distance measures $d$ between distributions. If observations $x$ are interpreted as outliers, the parameter of scientific interest may still be $\mu$ rather than $EQ_{\mu.\sigma^2,\epsilon,x}$, but the CLT will grant consistency of the sample mean for the latter, not the former. 

Cressie mentions that the biggest problems for the \textit{t}-test occur for violations of independence, however we are not aware of any literature examining of independence testing combined with the \textit{t}-test. 

Some work focuses just on the quality of the MS tests without specific reference to its effect on subsequent inference and combined procedures, see \cite{RazWah11,MenPal03,FarRog06,Keskin06}. This is of limited use, as it does not address which alternatives to normality are particularly problematic and for which a subsequent \textit{t}-test can still do well. \cite{ScHiWi06} and \cite{KeOtWi13} investigated normality tests regarding its use for subsequent inference without explicitly involving the later main test. Both advised against the Kolmogorov-Smirnov test. \cite{KeOtWi13} concluded that the Anderson-Darling test is the most effective one at detecting non-normality relevant to subsequent \textit{t}-testing, and they suggested that for deciding whether the MC test should be used, the MS test be carried out at a significance level larger than 0.05, for example 0.15 or 0.20, in order to increase the power, as all these tests may have difficulties to detect deviations that are problematic for the \textit{t}-test.

Another group of work examines running a \textit{t}-test conditionally on passing normality by a preliminary normality test. Most of these do not consider what happens if normality is rejected.  
\cite{EasAnd78} considered various distributions such as normal, uniform, exponential, two central and two non-central \textit{t}-distributions. They 
used both the Anderson-Darling and the Shapiro-Wilk normality tests. In the case that normality was passed, they compared the empirical distribution of the resulting \textit{t}-values to Student's \textit{t}-distribution. This worked well when the samples were drawn from the normal distribution (in which case the standardised residuals, on which normality tests can be run, are ancillary statistics independent of the estimated mean and variance). For symmetric non-normal distributions, the results were mixed, and for situations where the distributions were asymmetric, the distribution of the \textit{t}-values conditionally on not rejecting the normal assumption did not resemble a Student's \textit{t}-distribution, which they took as an argument against the practice of preliminary normality testing. As a result they favoured a nonparametric approach. 

\cite{SchNg06} investigated the conditional type 1 error rate of the one sample \textit{t}-test given that the sample has passed a test for normality for data from normal, uniform, exponential, and Cauchy populations. They conclude that the MS test makes matters worse in the sense that the Type I error rate is further away from the nominal 5\% (lower for the uniform and Cauchy, higher for the exponential) for data that pass the normality test than when the \textit{t}-test is used unconditionally, and this becomes worse for larger sample sizes. 
For the Cauchy distribution they also investigated running a Wilcoxon signed rank test as AU test conditionally on rejecting normality, which works worse than using the AU test unconditionally. \cite{RocKie11} came to similar conclusions using a somewhat different collection of MS tests and underlying distributions. 

\subsection{Tests of normality in the two-sample problem and one-way ANOVA}

For the two-sample problem, the Wilcoxon-Mann-Whitney (WMW) rank test is a popular alternative to the two-sample \textit{t}-test with (in the context of preliminary normality testing) mostly assumed equal variances. In principle most arguments and results from the one-sample problem apply here as well, with the additional complication that normality is assumed for both samples, and can be tested either by testing both samples separately, or by pooling residuals from the mean.  
As for the one-sample problem, there are also claims and results that the two-sample \textit{t}-test is rather robust to violations of the normality assumption (\cite{HsuFel69,RasGui04}), but also some evidence that this is sometimes not the case (see Example \ref{etwosam}), and that the WMW rank test can be superior and does not lose much power even if normality is fulfilled (\cite{NeaGra68}). \cite{FayPro10} presented a survey on comparing the two-sample \textit{t}-test with the WMW test (involving further options such as Welch's \textit{t}-test and a permutation \textit{t}-test for exploring its distribution under $H_0$), concluding that the WMW test is superior where underlying distributions are heavy tailed or contain a certain amount of outliers; it is well known that the power of the \textit{t}-test can break down under addition of a single outlier in the worst case, see \cite{HeSiPo90}. Although Fay and Proschan did not explicitly investigate deciding between \textit{t}- and WMW-test by normality testing, they advise against it, stating that normality tests tend to have little power for detecting distributions that cause problems for the \textit{t}-test. 

\cite{RoGoKi12} investigated by simulation combined procedures based on preliminary normality testing both for both samples separately, and pooled residuals using a Shapiro-Wilk test of normality. The MC test was the two sample \textit{t}-test, the AU test was the WMW test. Data were simulated from normal, exponential, and uniform distributions. In fact, for these distributions, the MC test was always better than the AU test, which makes a combined procedure superfluous; it reached acceptable performance characteristics, but inferior to the MC test. A truly heavy tailed distribution to challenge the MC test was not involved. 

\cite{Zimmerman11} achieved good simulation results with an alternative approach, namely running both the two-sample \textit{t}-test and the WMW test, choosing the two-sample \textit{t}-test in case the suitably standardised values of the test statistics are similar and the WMW test in case the p-values are very different. This seems to address the problem of detecting violations of normality better where it really matters. The tuning of this approach is somewhat less intuitive than for using a standard MS test. 

\cite{KeOtWi14} recommended preliminary normality testing in the multi-sample problem (one-way ANOVA), taking multiple testing issues into account. They investigated the power of their multi-sample misspecification test scheme, but not of the resulting combined procedure. \cite{LaAnMa16} investigated the use of a Shapiro-Wilk normality test (comparing various levels) to decide between running a standard ANOVA F-test and a nonparametric Kruskal-Wallis test under a number of normal and non-normal data generating processes. The corresponding combined procedure was recommended, because it displayed very slightly inflated type I error probabilities, but was close to the F-test where the F-test was better than Kruskal-Wallis, and close to Kruskal-Wallis where Kruskal-Wallis was better.

\subsection{Linear regression}\label{sreg}

In standard linear regression,
$$
y_i=\beta_0+\beta_1x_{1i}+\ldots+\beta_px_{pi}+e_i,\ i=1,\ldots,n,
$$
with response $Y=(y_1,\ldots,y_n)$ and explanatory variables $X_j=(x_{j1},\ldots,x_{jn}),\ j=1,\ldots,p$. $e_1,\ldots,e_n$ are in the simplest case assumed i.i.d. normally distributed with mean 0 and equal variances.  

The regression model selection problem is the problem to select a subset of a given set of explanatory variables $\{X_1,\ldots,X_p\}$. This can be framed as a model misspecification test problem, because a standard regression assumes that all variables that systematically influence the response variable are in the model. If it is of interest, as main test problem, to test $\beta_j=0$ for a specific $j$, the MS test would be a test of null hypotheses $\beta_k=0$ for one or more of the explanatory variables with $k\neq j$. The MC test would test  $\beta_j=0$ in a model with $X_k$ removed, and the AU test would test  $\beta_j=0$ in a model including $X_k$. This problem was mentioned as second example in \cite{Bancroft44}'s seminal paper on preliminary assumption testing. \cite{Spanos18} however argued that this is very different from MS testing in the earlier discussed settings, because if a model including $\beta_k$ is chosen based on a rejection of $\beta_k=0$ by what is interpreted as MS test, the conditionally estimated $\beta_k$ will be systematically too large in absolute value, and can through dependence on the estimated $\beta_j$ also be strongly dependent on the MC test.

Traditional model selection approaches such as forward selection and backward elimination are often based on such tests and have been analysed (and criticised) a lot in the literature. We will not review this literature here. There is sophisticated and innovative literature on post-selection inference in this problem. \cite{BeBrBuZhZh13} propose a procedure in which the main inference is adjusted for simultaneous testing taking into account all possible sub-models that could have been selected. \cite{Efron14} uses bootstrap methods to do inference that takes the model selection process into account. Both approaches could also involve other MS testing such as of normality, homoscedasticity, or linearity assumptions, as long as combined procedures are fully specified. \cite{ZhKhAs22} provide a recent survey of methods of this kind. For a critical perspective see \cite{LeePot05,LePoEw15}, noting particularly that asymptotic results regarding the distribution of post-selection statistics (i.e., results of combined procedures) will not be uniformly valid for finite samples. 
In econometrics, David Hendry and co-workers developed an automatic modeling system that involves MS testing and conditional subsequent testing with adjustments for decisions in the modeling process, see, e.g., \cite{HenDoo14}. They mentioned that their experience from experiments is that involving MS tests does not affect the final results much in case the model assumptions for the final procedure are fulfilled. However, to our knowledge these experiments are nowhere published. Earlier, some authors such as \cite{SalSen83} analysed the effect of preliminary variable selection testing on later conditional main testing.  

\cite{Godfrey88} listed a plethora of MS tests to test the various assumptions of linear regression. However, no systematic way to apply these tests was discussed. In fact, Godfrey noted that the literature left more questions open rather than answered. Some of these questions are: (i) the choice among different MS tests, (ii) whether to use nonparametric or parametric tests, (iii) what to do when any of the model assumptions are invalid as well as (iv) some potential problems with MS testing such as repeated use of data, multiple testing and pre-test bias. \cite{Godfrey96} concluded that efforts should be made to develop ``attractive'', useful and simple combined procedures as these were lacking at the time; to a large extent this still is the case. One suggestion was to use the Bonferroni correction for each test as ``\textit{the asymptotic dependence of test statistics is likely to be the rule, rather than the exception, and this will reduce the constructive value of individual checks for misspecification}''. 

\cite{GilGil93} reviewed the substantial amount of work done in econometrics regarding preliminary testing in regression up to that time, a limited amount of which is about MC and/or AU tests conditionally on MS tests. This involves pre-testing of a supposedly known fixed variance value, homoscedasticity, and independence against auto-correlation alternatives. The cited results are mixed. \cite{KinGil84} comment positively on a combined procedure in which absence of auto-correlation is tested first by a Durbin-Watson or \textit{t}-test. Conditionally on the result of that MS test, either a standard \textit{t}-test of a regression parameter was run (MC test), or a test based on an empirically generalised least squares estimator taking auto-correlation into account (AU test). In simulations the combined procedure performs similar to the MC test and better than the AU test in absence of auto-correlation, and similar to the AU test and better than the MC test in the presence of auto-correlation. Also here it is recommended to run the MS test at a level higher than the usual 5\%. Most related post-1993 work in econometrics seems to be on estimation after pre-testing, and regression model selection. \cite{OhtToy85} proposed a combined procedure for testing linear hypotheses in regression conditionally on testing for known variance. \cite{ToyOht86} tested the equality of different regressions conditionally on testing for equal variances. In both papers power gains for the combined procedure are reported, which are sometimes but not always accompanied by an increased type 1 error probability. 

While not defining a combined procedure or investigating specific MS tests, \cite{SchFin18} highlight biases resulting from applying transformations to the response variables in order to achieve better normality of residuals. They argue that inference from normality is often valid with large sample sizes even with non-normal residuals.

\subsection{Other problems}

Some specific problem-adapted 
combined procedures have been discussed in the literature. 

In a two-treatment, two-period cross-over trial, patients are randomly 
allocated either to one group that receives treatment A followed by treatment 
B, or to another group that receives the treatments in the reverse 
order. The straightforward analysis of such data could analyse within-patients
differences between the effects of the two treatments by a paired test (MC 
test). This 
requires the assumption that there is no ``carry-over'', i.e., no influence of
the earlier treatment on the effect of the later treatment. In case there
is carry-over, the somewhat wasteful analysis of the effect of the first 
treatment only for each patient is safer (AU test). 
\cite{Grizzle67} proposed a combined procedure that became well established for 
some time. It consists of computing a score for each patient that contrasts
the two treatment effects with the baseline values, and tests, e.g., using a 
two-sample {\textit t}-test, whether this is the same on average in 
both groups, 
corresponding to the absence of carry-over on average (MS test). 
\cite{Freeman89} analysed
this combined procedure analytically under a normal assumption and potential
existence of carry-over, comparing it to both the MC test and the AU test run
unconditionally. He observed that due to strong dependence between the MS test
and both the MC- and the AU-test, the combined procedure has more or less
strongly inflated type 1 errors whether there is carry-over or not.
Its power behaves typically for combined procedures, being better than the AU 
test but worse than the MC test in absence of carry-over and the other way round
in its presence. Overall Freeman advises against the use of this procedure. 

Similarly, \cite{Kahan13} advised against a combined procedure involving an MS test for the presence of interactions in a situation in which multiple treatments are assessed in a single trial, in order to decide whether a factorial analysis assuming no interactions should be used (MC test) rather than analysing the data assuming a four-arm trial with interactions (AU test). Often the MS test does not have enough power to detect interactions, but will also bias the analysis due to substantial dependence with the main tests. MC test and AU test on their own performed unsatisfactorily as well. \cite{Kahan13} ultimately recommended to run both MC and AU test and to only take results as meaningful in case the two tests coincide.

\cite{CamDea14} analysed combined procedures for the Cox proportional hazard 
model with covariates, in
which the assumption of a constant hazard rate over time is tested by a 
Grambsch and Therneau test (MS test). They considered a number of different approaches for the AU tests. Dependence between the MS test and the main tests produced inflated type 1 error probabilities with some approaches strongly and some mildly affected. The authors then applied a permutation adjustment to the combined procedure in case the covariate of interest is a treatment indicator. The distribution of the p-value of the combined procedure under the $H_0$ of no treatment effect can then be investigated permuting the treatment labels. This achieved unbiased type 1 error probabilities, with the power depending on the situation considered. This idea could also be applied in other situations where the null hypothesis is that a certain covariate, potentially a group indicator, has no effect. 

\subsection{More than one misspecification test}

\cite{RaKuMo11} assessed the statistical properties of a three-stage procedure including testing for normality and for homogeneity of the variances taking into account a number of different distributions, and ratios of the standard deviation. They considered three main statistical tests, the Student's $t$-test, the Welch's $t$-test and the WMW test. For the MS testing, they used the Kolmogorov-Smirnov test for testing normality and Levene's test for testing the homogeneity of the variances of the two generated samples (\cite{Levene60}). If normality was rejected by the Kolmogorov-Smirnov test, the WMW test was used. If normality was not rejected, the Levene's test was run and if homogeneity was rejected, the Welch's $t$-test was used and if homogeneity was not rejected, the standard $t$-test was used. 
Welch's $t$-test performed so well overall that the authors recommended its unconditional use, which is in line with recommendations by \cite{RasGui04} from investigations of the robustness of various tests against non-normality. All of the investigated distributions had finite kurtosis, meaning that the tails were not really heavy. 

\cite{McDrAl93} proposed a comprehensive strategy for linear regression involving a number of misspecification tests. They investigate its power against various types of misspecified models and recommend alternative models in case of misspecification, but do not investigate the performance of the resulting combined procedure.

\cite{Campbell21} investigated a combined procedure in which MS tests test for zero inflation and overdispersion in a Poisson regression model, with AU tests based on negative binomial, zero-inflated Poisson, and zero-inflated negative binomial models depending on the outcomes of two MS tests. Results turn out to be problematic at least in some situations for unconditional use of the MC or AU tests, but also for the combined procedures (model selection by AIC and BIC was also involved as an alternative to using MS tests). The combined procedure using MS tests was recommended for large sample sizes but may result in considerably inflated type 1 error probabilities for small samples.


\subsection{Discussion}\label{ssec5fd}

Although many authors have, in one way or another, investigated the effects of preliminary MS testing or later application of model-based procedures, there are severe limitations in the existing literature. Only few papers have compared  the performance of a fully specified combined procedure with unconditional uses of both the MC and the AU test. Some of these have only looked at type 1 error probabilities but not power (in many situations the behaviour of power is connected to the type 1 error, and improved power comes at the price of an increased type 1 error), and some have only looked at the situation in which the model assumption is in fact fulfilled. Some have studied setups in which either the unconditional MC or the AU test works well across the board, making a combined procedure superfluous, although it is widely acknowledged that situations in which either unconditional test can perform badly depending on the unknown data generating process do exist. In some other situations not only the combined procedure but also MC and AU test on their own perform unsatisfactorily.  
 
Reasons why authors advised against model checking in specific situations were: 
\begin{description}
\item[(a)] The MC test was better or at least not clearly worse than the AU test for all considered distributions in which the model assumptions of the MC test were not fulfilled (in which case the MC test can be used unconditionally). In some cases this has happened because authors did not include existing distributions in their simulations with which the MC test would have had bigger problems, e. g., severely 
heavy tailed distributions are missing from \cite{RocKie11}.
\item[(b)] The AU test was not clearly worse than the MC test where model assumptions of the MC test were fulfilled (in which case the AU test can be used unconditionally). E. g., the Welch and Sattertwaite tests work very well for equal variances in Section \ref{svar}.
\item[(c)] The MS test did not work well distinguishing situations in which the MC test was better from situations in which the AU test was better, possibly despite being good at testing just the formal model assumption.
\item[(d)] Due to dependence, the application of the MS test distorted the 
performance of the conditionally performed tests.
\item[(e)] The combined procedure was branded theoretically invalid, occasionally even when performing well in simulations, \cite{RoGoKi12}.
\item[(f)] In some work shortcomings of model checking before 
running the MC test are 
highlighted, but running the MC test without checking does not look overall 
convincing either, and both have not been compared with 
a full combined procedure or a specific AU test, e.g., \cite{EasAnd78,SchNg06}.
\cite{RoGoKi12} explicitly state that the full 
combined procedure performs better in their simulations than what could be 
expected from looking in an isolated manner at the effect of MS testing on each
single one of the main tests.
\end{description} 
For model checking to be worthwhile, (a)-(d) need to be avoided. (e) arguably is only an issue to the extent that the actual performance of the combined procedure is in fact bad, which seems to be covered by (a)-(d) already. 
Regarding (f), it is clear that general issues with MS testing
exist, but this 
does not necessarily mean that there is a better alternative.

A further recurring theme in the work investigating combined procedures is 
a recommendation to use a higher level for the MS test than the 
conventional 5\%, due to the difficulty of the involved MS tests to find
certain violations of the model assumptions that are critical for the later MC test.


Comparing a full combined procedure with unconditional use of
the MC test or the AU test, a typical pattern is
that under the model assumption
for the MC test, the MC test is best regarding power, and the combined 
procedure performs between the unconditional MC test and AU test, and if 
that model 
assumption is violated, the AU test is best, and the combined procedure is once
more between the MC test and the AU test, see Example \ref{etwosam}.
Such results can be interpreted
charitably for the combined procedure, protecting against the worst 
performances achieved by any of the main tests on their own. 
It seems to us that part of the criticism of the combined 
procedure is motivated by the fact that it does not do what some seem to expect
or hope it to do, namely to help making sure that model assumptions are 
fulfilled, and to otherwise leave performance characteristics untouched. 

A detailed
look at the results reveals that the combined procedures are almost
always competitive with at least one of the unconditional tests, and often
with them both (e.g., in \cite{LaAnMa16}).   
It is clear, though, that recommendations need to depend on
the specific problem and the specific tests involved. Results often also depend 
on in what way
exactly model assumptions of the MC test are violated, which is hard to know 
without some kind of data dependent reasoning.

\section{A positive result for combined procedures} \label{stheory}

The overall message from the literature does not seem very satisfactory. On the one hand, model assumptions are important and their violation can severely damage results. On the other hand, investigating the performance of combined procedures reveals severe limitations. 

In this section we present a setup and a result that makes us assess the impact of preliminary model testing somewhat more positively. A characteristic of the literature analysing combined procedures is that they compare the combined procedure with unconditional MC or AU tests both in situations where the model assumption of the MC test is fulfilled, or not fulfilled. However, they do not investigate a situation in which the MS test can do what it is supposed to do, namely to {\it distinguish} between these situations. 

We here model an idealised setup in which researchers in a certain field, say, analyse many data sets using the same tests based on certain model assumptions, and come across a certain percentage $0\le\lambda\le 1$ of situations in which the model assumptions are fulfilled, and a percentage of $1-\lambda$ situations in which the model assumptions are violated in a certain way.

Is such a setup relevant? 
Obviously it is not realistic that there are only two distributions possible, one of which fulfills the model assumptions of the MC test. We wanted to keep the setup simple, but of course one could look at mixtures of a wider range of distributions, even a continuous range (for example for ratios between group-wise variances). In any case, the setup is more flexible and one might say realistic than looking at $\lambda=0$ and $\lambda=1$ only, which is what has been done in the literature up to now. The aim here is to show in what sense the combined procedure can be better than both MC and AU tests applied unconditionally, and similar effects will likely be obtained in more complex setups. Also, model assumptions will of course never hold precisely, but having a certain probability that the model is wrong in a really problematic way whereas otherwise the nominal performance of the MC test is a good approximation does not seem unrealistic. 
The setup has a certain Bayesian flavor, but a researcher may want to use $\lambda$ as a simple device for decision support rather than being required to set up priors for $\lambda$ along with all other model parameters. 

Using the notation from Section \ref{scomb},
let $P_\theta$ be a distribution that fulfills the model assumptions of the MC test, and $Q\in M\setminus M_\Theta$ a distribution that violates these assumptions. For considerations of power, let the null hypothesis of the main test be violated, i.e., $\theta\not\in\Theta_0$ and $Q\not\in M^*$ (an analogous setup is possible for considerations of size). We may observe data from $P_\theta$ or from $Q$. Assume that a data set is with probability $\lambda\in[0,1]$ generated from $P_\theta$ and with probability $1-\lambda$ from $Q$ (we stress that as opposed to standard mixture models, $\lambda$ governs the distribution of the whole data set, not every single observation independently). The cases $\lambda=0$ and $\lambda=1$ are those that have been treated in the literature, but only if $\lambda\in(0,1)$ the ability of the MS test to inform the researcher whether the data are more likely from $P_\theta$ or from $Q$ is actually required.

\begin{figure}[t]
\caption{Power of combined procedure, MC (Welch's two-sample \textit{t}-test), and AU test (WMW-test) across different $\lambda$s. The setup is from Example \ref{etwosam}.}
\begin{center}
\includegraphics[width=0.7\textwidth]{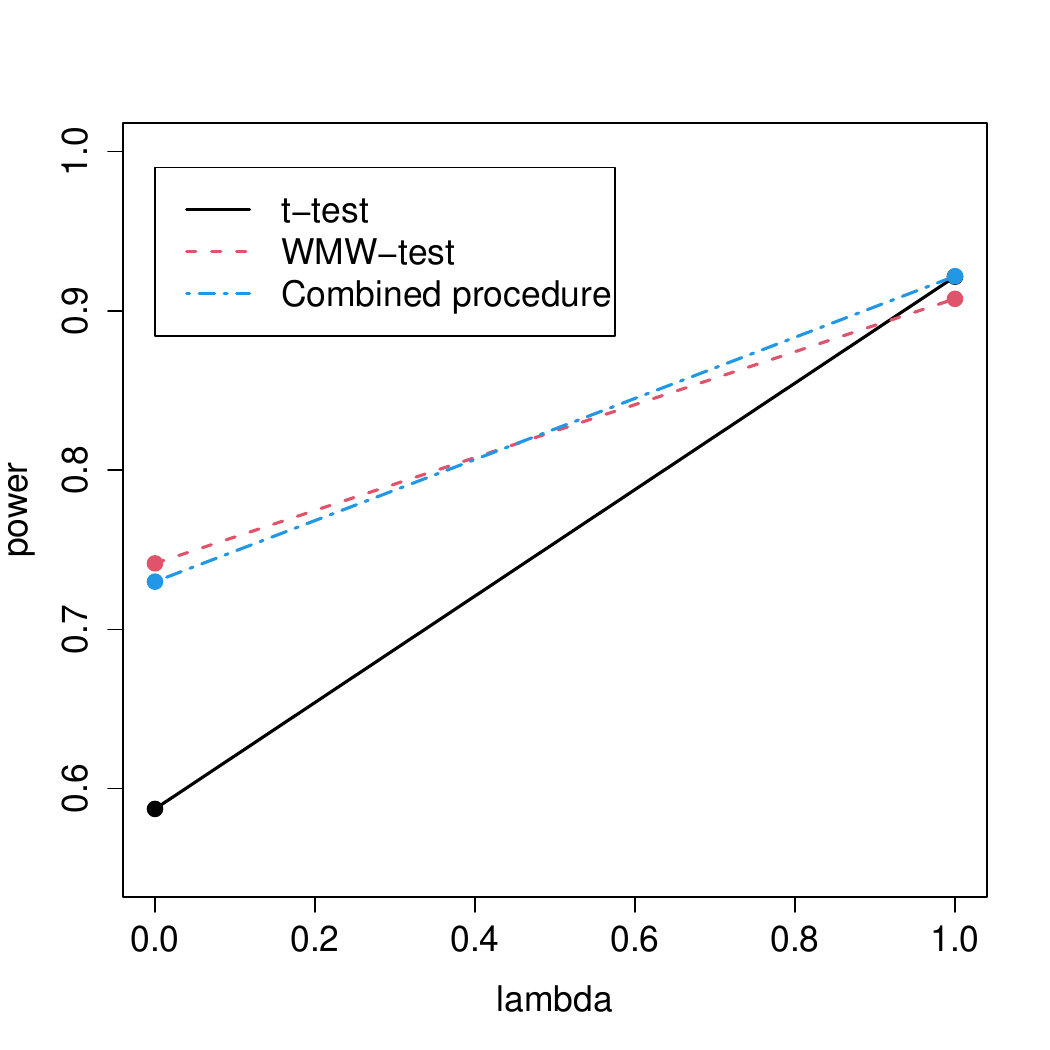}
\end{center}
\label{fig:simueg}
\end{figure}

Figure \ref{fig:simueg} is based on Example \ref{etwosam} with $P_{\theta}$ being the normal and $Q$ the $t_3$-distribution. In this situation, for $\lambda=0$ ($t_3$; model assumption violated), the AU test (WMW) is best and the MC test ({\it t} test) is worst. For $\lambda=1$, both MC test and combined procedure reach about the same power, and the AU test is worst. The powers of all three tests are linear functions of $\lambda$, and the consequence is that the combined procedure performs better than both unconditional tests over the best part of the range of $\lambda$, and never much worse than the best. The latter can be seen in many setups of this kind. This good result is somewhat contrasted by a slightly inflated type 1 error probability of the combined procedure though, see Example \ref{etwosam}.

On top of the notation from Section \ref{scomb}, $P_\lambda$ stands for distribution of the overall two step experiment, i.e., first selecting either $\tilde{P}=P_\theta$ or $\tilde{P}=Q$ with probabilities $\lambda,\ 1-\lambda$ respectively, and then generating a data set $z$ from $\tilde{P}$. The events of rejection of the respective $H_0$ are denoted $R_{MS}=\{\Phi_{MS}(z)=1\}$, $R_{MC}=\{\Phi_{MC}(z)=1\}$, $R_{AU}=\{\Phi_{AU}(z)=1\}$, $R_{C}=\{\Phi_{C}(z)=1\}$, where $C$ refers to the combined procedure. Here are some assumptions:
\begin{description}
	\item[(I)] $\Delta_{\theta} = P_\theta(R_{MC}) - P_\theta(R_{AU})>0$,
	\item[(II)] $\Delta_Q = Q(R_{AU}) - Q(R_{MC})>0$,
        \item[(III)] $\alpha^*_{MS}=Q(R_{MS})>\alpha_{MS}=P_\theta(R_{MS})$,
        \item[(IV)] Both $R_{MC}$ and $R_{AU}$ are independent of $R_{MS}$ under both $P_\theta$ and $Q$.
\end{description} 
Keep in mind that this is about power, i.e., we take the $H_0$ of the main test as violated for both $P_\theta$ and $Q$.
Assumption (I) means that the MC test has better power under $P_\theta$, (II) means that the AU test has better power under $Q$. Assumption (III) means that the MS test has some use, i.e., it has a certain (possibly weak) ability to distinguish between $P_\theta$ and $Q$. All these are essential requirements for preliminary model assumption testing to make sense. Assumption (IV) though is very restrictive. It asks that rejection of the main null hypothesis by both main tests is independent of the decision made by the MS test. This is unrealistic in most situations, but it can be relaxed, as is done in Corollary \ref{cor1} below, which requires only approximate independence.
As emphasised earlier, approximate independence of the MS test and the main tests has also been found in other literature to be an important desirable feature of a combined test, and it should not surprise that a condition of this kind is required.  

The following Lemma states that the combined procedure has better power than both the MC test and the AU test for at least some $\lambda$. Although this in itself is not a particularly strong result, in many situations, according to our experience, the range of $\lambda$ for which this holds is quite large. Furthermore the result concerns general models and choices of tests, whereas to our knowledge everything that already exists in the literature is for specific choices.
  
Despite the somewhat restrictive set of assumptions, none of the involved tests and distributions is actually specified, so that the Lemma (at least with a relaxed version of (IV)) applies to a very wide range of problems.   
\begin{lemma}\label{lemma1} Assuming (I)-(IV), $\exists \lambda \in (0,1)$ such that both $P_\lambda(R_C)>P_\lambda(R_{MC})$ and $P_\lambda(R_C)>P_\lambda(R_{AU})$.
\end{lemma}
\begin{proof} Obviously,
\begin{align*}
P_\lambda(R_{MC})&= \lambda P_\theta(R_{MC})+(1-\lambda)Q(R_{MC}), \\
P_\lambda(R_{AU})&= \lambda P_\theta(R_{AU})+(1-\lambda)Q(R_{AU}). \\
\end{align*}
By (I), for $\lambda=1:\ P_\lambda(R_{MC})>P_\lambda(R_{AU})$ and, by (II), 
for $\lambda=0:\ P_\lambda(R_{AU})>P_\lambda(R_{MC})$. As $P_\lambda(R_{MC})$ and
$P_\lambda(R_{AU})$ are linear functions of $\lambda$, there must be 
$\lambda^*\in(0,1)$ so that $P_{\lambda^*}(R_{AU})=P_{\lambda^*}(R_{MC})$.
Obtain
\begin{align*}
P_{\lambda^*}(R_{MC})& =P_{\lambda^*}(R_{AU}) \Leftrightarrow\\
\lambda^* P_\theta(R_{MC})+(1-\lambda^*)Q(R_{MC})&= \lambda^* P_\theta(R_{AU})+(1-\lambda^*)Q(R_{AU})\Leftrightarrow\\
\lambda^*(\Delta_\theta+\Delta_Q)=\Delta_Q \Leftrightarrow\\
\lambda^*=\frac{\Delta_Q}{\Delta_\theta+\Delta_Q}.
\end{align*}
This yields, by help of (IV), 
\begin{align*}
P_{\lambda^*}(R_C) &= \lambda^* P_\theta(R_C)+(1-\lambda^*)Q(R_C)\\
&=\lambda^*\left[\alpha_{MS}P_\theta(R_{AU}|R_{MS})+(1-\alpha_{MS})P_\theta(R_{MC}|R_{MS}^c)\right]\\
&+(1-\lambda^*)\left[\alpha_{MS}^*Q(R_{AU}|R_{MS})+(1-\alpha_{MS}^*)Q(R_{MC}|R_{MS}^c)\right]\\
&=\lambda^*\left[\alpha_{MS}P_\theta(R_{AU})+(1-\alpha_{MS})P_\theta(R_{MC})\right]\\
&+(1-\lambda^*)\left[\alpha_{MS}^*Q(R_{AU})+(1-\alpha_{MS}^*)Q(R_{MC})\right]\\
&=\frac{\Delta_Q}{\Delta_\theta+\Delta_Q}\left[-\alpha_{MS}\Delta_\theta-\alpha_{MS}^*\Delta_Q\right]+\alpha_{MS}^*\Delta_Q+P_{\lambda^*}(R_{MC})\\
&=\Delta_Q\left[\frac{-\alpha_{MS}\Delta_\theta - \alpha^*_{MS}\Delta_Q + \alpha^*_{MS}\Delta_\theta + \alpha^*_{MS}\Delta_Q}{\Delta_\theta+\Delta_Q} \right] + P_{\lambda^*}(R_{MC})\\
&= \frac{\Delta_\theta\Delta_Q}{\Delta_\theta+\Delta_Q}\left[\alpha^*_{MS} - \alpha_{MS}\right] + P_{\lambda^*}(R_{MC})\\
&=  \frac{\Delta_\theta\Delta_Q}{\Delta_\theta+\Delta_Q}\left[\alpha^*_{MS} - \alpha_{MS}\right] + P_{\lambda^*}(R_{AU}).
\end{align*}
$\frac{\Delta_\theta\Delta_Q}{\Delta_\theta+\Delta_Q}\left[\alpha^*_{MS} - \alpha_{MS}\right]$ is larger than zero by (I)-(III), so $P_{\lambda^*}(R_C)$ is larger than both $P_{\lambda^*}(R_{MC})$ and $P_{\lambda^*}(R_{AU})$.
\end{proof}
\begin{corollary} \label{cor1}
Assuming (I)-(III), Lemma \ref{lemma1} still holds if there is a small enough $\delta>0$ (dependent on the involved probabilities) so that $|P_\theta(R_{MC}|R_{MS})-P_\theta(R_{MC}|R_{MS}^c)|$, $|P_\theta(R_{AU}|R_{MS})-P_\theta(R_{AU}|R_{MS}^c)|$,  $|Q(R_{MC}|R_{MS})-Q(R_{MC}|R_{MS}^c)|$, and $|Q(R_{AU}|R_{MS})-Q(R_{AU}|R_{MS}^c)|$ are all smaller than $\delta$.
\end{corollary}
\begin{proof} This follows from the continuity of the sums and products in the proof of Lemma \ref{lemma1}.
\end{proof}
\begin{example} For given $\lambda$, error probabilities of all procedures are just convex combinations of the error probabilities for $\lambda=0$ and $\lambda=1$. In a given situation with sample sizes known in advance, simulations as in Example \ref{etwosam} can be used, together with an assumed value of $\lambda$, in order to compare the procedures and to preselect an ``optimal'' one. This easily extends to more than two possible distributions. Assume in the situation of Example \ref{etwosam} that a researcher commits themselves to expect an approximate normal distribution with probability $\frac{1}{2}$, an approximate $t_3$-distribution with probability $\frac{1}{4}$ and an approximate skew normal distribution with probability $\frac{1}{4}$ (if measurements can take negative values, the exponential does not seem to be a valid option; on the other hand, in situations in which only positive values can be taken, one could simulate, e.g., zero-truncated normal distributions, and even discreteness of data, e.g., knowledge that they all come with at most one digit after the decimal point, can be simulated). Obviously this does not mean that it is required that no other situation can occur, it just examines a ``point of orientation'', and is surely more informative than a standard consideration of the nominal model assumption only. This yields the type 2 (type 1 in brackets) error probabilities .1641 (.049) for Welch's {\textit t}-test, 0.1303 (.048) for WMW, 0.1250 (0.052) for the combined procedure, and .1619 (0.048) for the permutation test, using the values in Table \ref{ttwosam1}. The combined procedure does best according to power here, and for a good range of choices of ``prior probabilities'', although anti-conservativity is a concern. 
\end{example}

\section{Conclusion} \label{sconc}

Given that statisticians often emphasise that statistical inference relies on model assumptions, and that these need to be checked, the literature investigating this practice is surprisingly critical. 
In some setups either running a less constrained test or running the model-based test without preliminary testing have been found superior to the combined procedure involving preliminary MS testing. This is in contrast to a fairly general view among statisticians that model assumptions should be checked. The existence of situations in which performance characteristics rely strongly on whether model assumptions are fulfilled or not has been acknowledged also by authors that were more critical of preliminary testing, and therefore there is certainly a role for model checking. There is however little elaboration of its benefits in the literature. A key contribution of the present work is the investigation of general combined procedures in a setup in which both distributions fulfilling and violating model assumptions can occur. This is more favourable for combined procedures than just looking at either fulfilled or violated model assumptions in isolation.
 

We believe that much literature gives a somewhat too pessimistic assessment of combined procedures involving MS testing (some reasons are listed in Section \ref{ssec5fd}), and that model checking (and drawing consequences from the result) is more useful than some of the literature suggests. The fact that preliminary assumption checking technically violates the assumptions it is meant to secure is probably assessed more negatively from the position that models can and should be ``true'', whereas it may be a rather mild problem if it is acknowledged that model assumptions, while providing ideal and potentially optimal conditions for the application of model-based procedures, are not necessary conditions for their use. 

On the other hand we believe that model checking is widely misconceived; its role should not be to make sure that the formal assumptions hold (be it approximately), but rather to find the particular violations of model assumptions that are most problematic in terms of level and power. With a better understanding of this point, research can be directed at finding MS tests (or more general decision rules) that fulfill the latter aim better.  

Lemma \ref{lemma1} gives an idea of the required ingredients for successful model checking, i.e., conditions for the combined procedure to be superior to both the MC and the AU test. In order to put this into practice, the researcher should have at least a rough idea about what kinds of deviations from the model assumptions of the MC test may happen, although one may also use ``worst cases'' (such as distributions with non-existing variances for \textit{t}-tests) as a starting point. Call $\{P_\theta\}$ the family of distributions that fulfill the model assumptions of the MC test, and $Q$ a possible distribution that violates them; one can also involve different options for $Q$. If available, prior information about in what way model assumptions may be violated should be used, as also emphasised by \cite{Tijmstra18} as key ingredient for model checking.   
\begin{description}
\item[(a)] The MC test should be clearly better than the AU test if its model assumptions are fulfilled (otherwise the unconditional AU test can be used without much performance loss).
\item[(b)] The AU test should be clearly better than the MC test for $Q$ (otherwise the unconditional MC test can be used without much performance loss).
\item[(c)] The MS test should be good at distinguishing $\{P_\theta\}$ from $Q$.
\item[(d)] The MS test $\Phi_{MS}$ should be approximately independent of both $\Phi_{MC}$ and $\Phi_{AU}$ under $\{P_\theta\}$ and $Q$. 
\end{description}
In practice it is not known what $Q$ will be encountered, but given the unsatisfactory state of the art, developing combined procedures fulfilling (a)-(d) based on realistic choices of $Q$ seems a promising approach to improve matters. Following the literature, in some situations it will be advantageous to run the MS test at a higher level than usual. A reviewer suggested that combined procedures could be generalised by using the result of the MS test to estimate weights for the MC and AU test to combine their results in a continuous manner.  

Considering informal (visual) model checking, issues (a) and (b) are not different from formal combined procedures, although the visual display may help to pick a suitable AU test (be it implicitly by formulating a model that does not require a rejected assumption). An expert data analyst may do better based on suitable plots than existing formal procedures regarding (c); many users will probably do worse (see \cite{HoKiJo12} for a study investigating misconceptions and lack of knowledge about model checking among empirical researchers). (d) may be plausible if displays are used in which the parameters tested by the MC and AU test such as location or regression parameters do not have a visible impact, such as residual plots, although there is a danger of this being critically violated in case the AU test is chosen based on what is seen in the plots.  


We believe that the approach of Lemma \ref{lemma1} considering a random draw of either fulfilled and violated model assumptions could also help in more complex situations, for example concerning different assumption violations, more than one MS test, and more than two main tests. 




\begin{thebibliography}{}
\bibitem[Pearson, 1900]{Pearson00}
Pearson, K. (1900).
\newblock On the criterion that a given system of deviations from the probable
  in the case of a correlated system of variables is such that it can be
  reasonably supposed to have arisen from random sampling.
\newblock {\em Philosophical Magazine}, 5:157--175.
\end{thebibliography}

\bibliographystyle{chicago}

\end{document}